\documentclass[namedreferences]{solarphysics}
\usepackage[hyperref,optionalrh]{spr-sola-addons} % For Solar Physics 
\usepackage{graphicx}        % For eps figures, newer & more powerfull
\usepackage{color}    
\usepackage{soul}% For color text: \color command
\renewcommand{\vec}[1]{{\mathbfit #1}}

\chardef\us=`\_

\begin{document}

\begin{article}
\begin{opening}

\title{On the Magnetic Reconnection and its Properties during a Flare Using a Magnetohydrodynamics Simulation}

\author[addressref={aff1,aff2},email={s.sangeetanayak93@gmail.com}]{\inits{S.}\fnm{Sushree S.}~\lnm{Nayak}\orcid{0000-0002-4241-627X}}
\author[addressref={aff1,aff3}]{\inits{}\fnm{Qiang}~\lnm{Hu}\orcid{0000-0002-7570-2301}}
\author[addressref={aff1,aff4}]{\inits{}\fnm{Wen}~\lnm{He}\orcid{0000-0001-8749-1022}}
\author[addressref={aff5}]{\inits{}\fnm{Sanjay}~\lnm{Kumar}\orcid{0000-0003-4578-6572}}
\author[addressref={aff6}]{\inits{}\fnm{Ramit}~\lnm{Bhattacharyya}\orcid{0000-0003-4522-5070}}

\address[id=aff1]{Center for Space Plasma and Aeronomic Research, The University of Alabama in Huntsville, Huntsville, AL 35899, USA.}
\address[id=aff2]{Indian Institute of Astrophysics, Koramangala, Bengaluru, 560034, India}
\address[id=aff3]{Department of Space Science, The University of Alabama in Huntsville, Huntsville, AL 35899, USA.}
\address[id=aff4]{New Jersey Institute of Technology, University Heights, Newark, NJ 07102-1982, USA}
\address[id=aff5]{Department of Physics, Patna University, Patna-800005.}
\address[id=aff6]{Udaipur Solar Observatory, Physical Research Laboratory, Dewali, Bari Road, Udaipur, 313001, India}
\runningauthor{Nayak et al.}
\runningtitle{\textit{Flare ribbon dynamics from simulation}}

\begin{abstract}
 We study the magnetic reconnection during a flare by investigating flare ribbon dynamics using observations and data-constrained magnetohydrodynamics (MHD) simulation. In particular, we estimate the reconnection flux and the reconnection flux rates using flare ribbons of an M1.1 flare hosted by the active region 12184 utilizing the technique developed by Qiu et al. (2002, ApJ, 565, 1335). The reconnection flux and corresponding flux rates are found to be  $10^{20}$ Mx and $10^{18}$ Mx s$^{-1}$ respectively. To understand the flare onset and the origin of flare ribbons, we perform an MHD simulation initiated by the non-force-free-field extrapolation. Importantly, the extrapolated configuration identifies a three-dimensional (3D) magnetic neutral point and a flux rope in the flaring region, which is crucial to the flaring activity. The reconnection initiates at the null point and, subsequently the flux rope rises and appears to reconnect there, which is favorable for the eruption of the filament. The surrounding field lines also seem to take part in the null point reconnection. In later stage, a current sheet is formed below the null point ensuing a secondary reconnection near an X-type topology, further contributing to the energy release process in the flare. We trace the footpoint evolution of the field lines lying over the flare ribbons and find a significant similarity between the observed flare ribbons and the evolution of footpoints computed from the MHD simulation. We estimated induced electric field during the flare and found it to be $\approx$ .52 V cm$^{-1}$, a slight less value, as per many past literatures. Additional findings are the enhancement of vertical current density near the flaring ribbons, a signature of successive reconnections near the null point. Overall, the present work contributes to the understanding of the ribbon formation in a flaring process and the involved magnetic reconnection. 
 
\end{abstract}
\end{opening}
%-------------------------------------------------

\section{Introduction}
     \label{intro} 

Magnetic reconnection is believed to be one of the underlying mechanisms of many solar transients viz. flares, coronal mass ejections (CMEs), and jets {\citep{priest-new}}. Magnetic reconnection involves the rearrangement of the magnetic-field 
topology along with a conversion of the magnetic energy into kinetic energy, heat, and acceleration of charged particles from the reconnection site and also upon the action of termination shocks {\citep{priest-new,2015Sci...350.1238C}}. With the advent of new instruments for observations and improved numerical simulations, attempts have been made to explore the underlying physics of reconnection in the solar atmosphere {\citep{priest-forbes1989, priest-new}}. However, a full understanding of reconnection physics is not yet achieved {\citep{priest-new}}, especially in three-dimensional (3D) settings.

Relevantly, the standard flare model or the CSHKP model \citep{1964NASSP..50..451C, 1968IAUS...35..471S, 1974SoPh...34..323H, 1976SoPh...50...85K} provides a physical scenario for the eruptive flares. According to this model, a filament or magnetic flux rope escapes the over-arching magnetic-field lines while creating a current sheet below it, ensuing reconnection there. As a result, the accelerated charged particles are produced from the reconnection site and under the effect of termination shock and move along the field lines toward the denser plasma at lower heights --- heating the dense atmosphere to create quintessential flare ribbons along with the formation of the post-flare loops. The reconnecting loops appear to move away from the polarity inversion line (PIL) while reconnecting site ascends in time.

In magnetic reconnection, one of the important parameters is the reconnection rate, which defines the amount of magnetic flux infused into the diffusion region per unit of time with an inflow ($v_{\rm{in}}$), given by $v_{\rm{in}}B$ \citep{2005ApJ...632.1184I}. In non-dimensional form, the rate is expressed as $M = v_{\rm{in}}/v_{A}$, where $v_{A}$ is the Alfv\'en velocity. From the Sweet-Parker model \citep{1957JGR....62..509P, 1958IAUS....6..123S}, the reconnection rate is $M_{A} = S^{-1/2} \approx 10^{-7}$,  where $S = \frac{Lv_{A}}{\eta}$ is the Lundquist number with $\eta$ being the magnetic diffusivity considering Spitzer resistivity \citep{1956pfig.book.....S} for solar corona.  \citet{1964NASSP..50..425P} estimated the rate to be  $M_{A} = \pi/8 \log{S} \approx 0.01-0.1$ in two-dimensional (2D) models considering the slow sock effects. \citet{1986PhFl...29.1520B} through a numerical approach agreed to the Sweet-Parker model result, while \citet{1986PhFl...29.3659U, 1989JGR....94.8805S, 1994ApJ...436L.197Y} have proven the rate to be in favor of Petschek-type fast reconnection with an anomalous resistivity in their magnetohydrodynamics (MHD) simulations. 

On the other hand, estimation of the reconnection rate from observations is challenging due to the difficulty in measuring the coronal magnetic field ($B_{\rm{corona}}$) and the inflow velocity there. Noticeably, direct plasma flows were first seen in limb flares observed in EUV images of SOHO/EIT \citep{2001ApJ...546L..69Y, 2006ApJ...637.1122N}. Noticing the successive production of flare ribbons in the chromosphere is an indication of the successive reconnection at the coronal height, and flare ribbons in the chromosphere are then used to indirectly measure the reconnection rate \citep{2000JASTP..62.1499F, 2004SoPh..222..115L}. In this approach, the reconnection rate in a 2D geometry is equal to the electric field in the reconnecting current sheet given by $E_{\rm{cs}} = v_{\rm{in}} B_{\rm{corona}} = v_{\rm{foot}} B_{\rm{foot}}$, where  $B_{\rm{foot}}$ is the magnetic field strength at the footpoint of the loops and $v_{\rm{foot}}$ is the footpoints separation velocity on the ribbons.  \citet{2002ApJ...566..528I, 2005ApJ...632.1184I} have adapted this relation to obtain the reconnection rate in a flare.  A detailed discussion about the measurement method of reconnection rate using the flare ribbons can be found in \citet{2007ApJ...659..758Q,2010ApJ...725..319Q}. The method has been successfully and extensively utilized by \citet{2017ApJ...845...49K} to understand the correlation between the flare ribbon fluxes and the peak X-ray flux for a large number of flares. Using the same method, a recent study by \citet{2022SoPh..297...80Q} infers the role of magnetic shear in influencing the reconnection rate.

Nevertheless, many flares are atypical by not falling into the standard flare model scenario. \citet{2014ApJ...791L..13L} in their study of a large set of confined flares reported the role of slipping motion of flares and ribbons near the quasi-separatrix layers \citep[QSLs:][]{1995JGR...10023443P}. Notably, a QSL is a potential site of magnetic reconnection, where a sharp change in magnetic field connectivity is found.
Further in their subsequent work, \citet{2015ApJ...804L...8L} reported the quasi-periodic slipping motion of flare loops during an X-class flare. Again, \citet{2022ApJ...933..191D} showed the role of slipping magnetic reconnection in a flare and later the slippage of the flux rope. Afterward, many observations were reported for other types of flare ribbons like circular flare ribbons, and not-so-symmetric parallel flare ribbons in many instances by \citet{2014ApJ...784..144D, 2015ApJ...807...72L, 2016ApJ...823...41D, prasad+2018apj, 2019ApJ...883...96Z, 2019ApJ...887..118C, 2020ApJ...899...34L}. Hence, the variety in flare ribbon structures makes it difficult to explain the different properties such as the reconnection rate of the underlying reconnection process.

Against the above backdrop, in this paper, we have investigated the role of magnetic reconnection in the formation of non-parallel ribbons during an M-class flare on 23 May 2021 hosted by a complex active region in terms of the distribution of magnetic fluxes. For this purpose, we have utilized the extrapolation model known as the non-force-free-field (NFFF) extrapolation \citep{2006GeoRL..3315106H, 2008ApJ...679..848H, 2010JASTP..72..219H} to construct the coronal magnetic field configuration. The extrapolated configuration suggests the presence of a 3D null as well as a flux rope in the flaring region. To explore the dynamical evolution of the flare, we perform a data-constrained MHD simulation initialized with the NFFF extrapolated field using the numerical model EULAG-MHD \citep{smolarkiewicz&charbonneau2013jcoph}. 
In the numerical model, the reconnections are achieved by the intermittent and adaptive numerical diffusion to the locations of under-resolved scales in the computational volume while fulfilling the flux-freezing condition elsewhere {\citep{2016PhPl...23d4501K, 2017PhPl...24h2902K}}. The simulated dynamics documents the onset of reconnections at the 3D null that trigger the flaring activity and form the flare ribbons. These reconnections continue in time and allow the flux rope to reconnect at the null, which potentially leads to the filament eruption.  Additionally, in the simulation, we have calculated the reconnection rate and reconnection flux rate near the flare ribbons. Notably, using the method by {\citet{2002ApJ...565.1335Q}}, we have also estimated the reconnection flux and reconnection flux rate observationally, which $\approx$ $10^{20}$ Mx and $10^{18}$ Mx s$^{-1}$ respectively. The values are one order less than the ones calculated in the simulation. Additional estimations on electric fields near the null point region is carried out along with current density on the ribbons. The derived reconnection rate from the simulation is found to be slightly less than the values obtained in earlier works. 

The rest of the paper is arranged as follows. Section \ref{obs} outlines the progress of the flare from observations. Section \ref{initial} 
describes the extrapolation model, and presents the initial extrapolated magnetic field. Next, Section \ref{numeric} describes the MHD model and the simulation set-up for the study. Section \ref{result} discusses the results of the simulation. Lastly, Section \ref{conclusion} summarizes the work.

\section{The M-class Flare in AR 12824}
\label{obs}
\begin{figure}
\centering
\includegraphics[width=\textwidth]{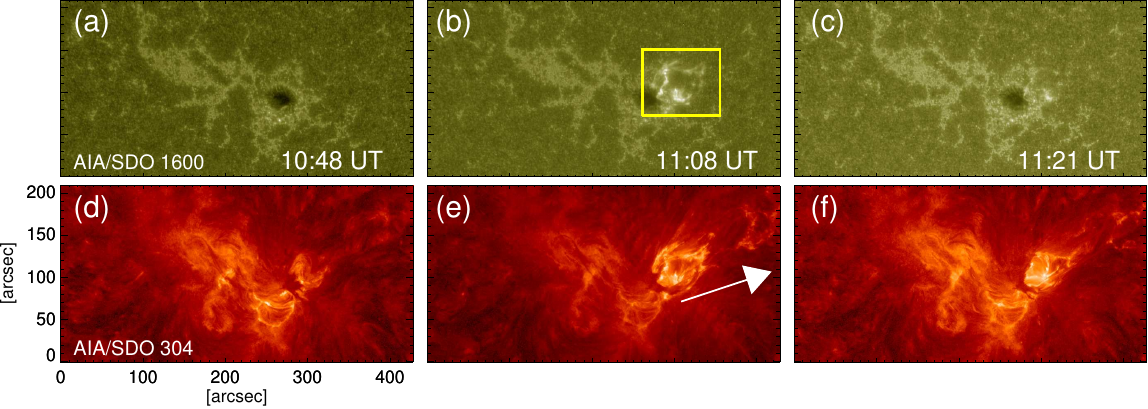}
\caption{The history of the M-class flare: \textit{Panels (a)-(c)} represent the evolution of the flare ribbons in 1600 {\AA} of AIA/SDO. \textit{Panel (b)} depicts the growth of the ribbons (marked by \textit{yellow box}) at the peak time of the flare. \textit{Panels (d)-(f)} show the evolution of the flare in 304 {\AA} of AIA/SDO and \textit{white arrow
}in \textit{panel (e)} shows the direction of the motion of the eruption. Important is the appearance of non-parallel flare ribbons. }
\label{aia}
\end{figure}

We have studied the M1.1 flare on 23 May 2021 in the active region NOAA 12824, which was initiated at  10:48 UT and lasted for $\approx 40$ minutes (\url{hinode.isee.nagoya-u.ac.jp/flare_catalogue/}). The progress of the flare is depicted in 1600 {\AA} (panels (a) to (c)) and 304 {\AA} (panels (d) to (f)) of Atmospheric Imaging Assembly \citep[AIA:][]{Lemen2012} onboard the Solar Dynamics Observatory \citep[SDO:][]{2012SoPh..275..207S,2012SoPh..275....3P} in Figure \ref{aia}. Panels (a) and (d) show the flaring region at 10:48 UT which is $\approx 18$ minutes before the peak time of the flare. Notable is the growth of the ribbons in panel (b), marked by yellow color box and then the decay phase comes, plotted in panel (c). The ribbons are not parallel and have a non-standard morphology. The 304 {\AA} observations of the flare (panels (d) to (f)) document the eruptive motion of the filament (marked by a white arrow in panel (e)) during the flare.

%--------Extrapolation----------------
\section{The Non-Force-Free-Field Extrapolation Model}
\label{initial}
To construct the magnetic configuration of the flaring region, we extrapolate the photospheric magnetic field of the active region utilizing the NFFF model. The model is based on the minimum dissipation rate principle (MDR) where the relaxed state of a two-fluid plasma system is achieved by minimizing the total dissipation rate keeping the generalized helicity constant \citep{2004PhPl...11.5615B, 2007SoPh..240...63B}. The relaxed state is a non-force-free state and supports a non-zero Lorentz force, which is used to self-consistently drive the plasma in the presented MHD simulation.  Briefly, The extrapolation solves an inhomogeneous double-curl
Beltrami equation for the magnetic field $\vec{B}$ (\citep{2004PhPl...11.5615B, 2007SoPh..240...63B}, and references therein),

\begin{equation}
\label{bext}
 	\nabla \times \nabla \times \vec{B} + a_{1} \nabla \times \vec{B} + b_{1} \vec{B} = \nabla \psi,
\end{equation}
\noindent
where $a_{1}$ and $b_{1}$ are constants. The solenoidality of $\vec{B}$ enforces
the scalar function $\psi$ to obey Laplace's equation. The solution to Equation \ref{bext}, after solving a two-fluid steady state 
\citep{1998PhRvL..81.4863M}, is a superposition of three subfields as, 
\begin{equation}
 	\vec{B} = \vec{B}_{3} + \sum_{i=1,2} \vec{B}_{i},
\end{equation}
\noindent where $\vec{B}_{3} = \nabla \psi$ is a potential field and the other two $\vec{B}$s are linear force-free fields having two different $\alpha$ satisfying the $\nabla \times \vec{B}_{i} = \alpha_{i} \vec{B}_{i},$ and related by $a_{1} = - (\alpha_{1}+\alpha_{2})$ and $b_{1} = \alpha_{1}\alpha_{2}$. Following \citet{2010JASTP..72..219H}, an optimal pair of $\alpha$ is
computed for $\vec{B}_{3} = 0$ by minimizing the average normalized
deviation of the magnetogram transverse field $B_{t}$ from its
extrapolated value $b_{t}$, given by
\begin{equation}
	E_{n} =  \left( \sum_{i=1}^M |\vec{B}_{t,i} - \vec{b}_{t,i}| \times |\vec{B}_{t,i}|\right) \bigg/\left(\sum_{i=1}^M |\vec{B}_{t,i}|^2 \right),
\end{equation}
\noindent where $M = N \times N$ is the total number of grid points on the
transverse plane. Additional minimization of $E_{n}$ is done by
using $\vec{B}_{3} = \nabla \psi$ as a corrector field for the obtained pair of $\alpha$. Details of the NFFF extrapolation models can be found in \citet{2006GeoRL..3315106H, 2008ApJ...679..848H, 2010JASTP..72..219H}.

\subsection{The NFFF Extrapolated Topology of AR 12884} \label{sec:extfield}    
\begin{figure}
\centering
   \includegraphics[width=.95\linewidth]{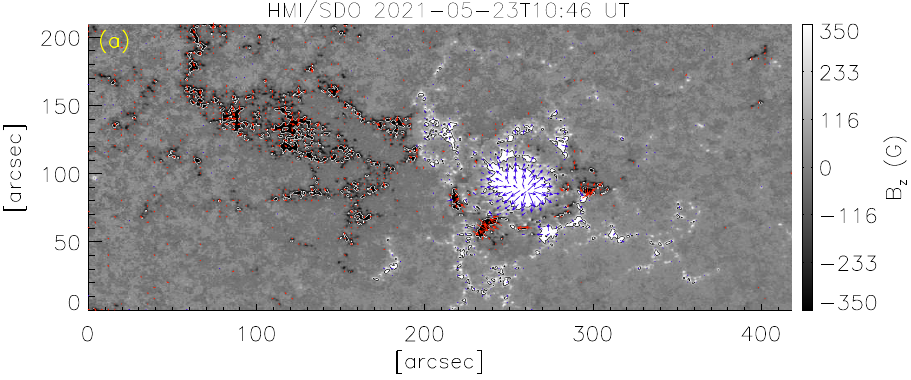}
    \includegraphics[width=.95\linewidth]{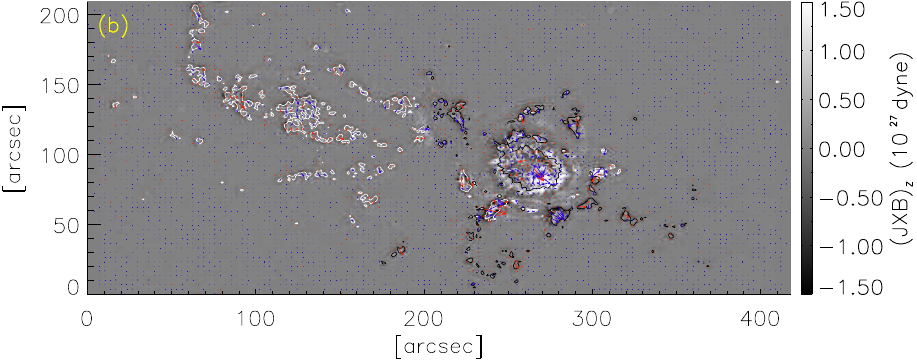}
\caption{\textit{Top}: vector plot of magnetic field, where \textit{red} and \textit{blue arrows} denote the transverse components and the \textit{grayscale} denotes the longitudinal component of $\vec{B}$ of HMI/SDO in \textit{panel (a)}. \textit{Panel (b)} shows the vector plot of the Lorentz force on the bottom boundary ($z=0$) derived from the extrapolated 3D magnetic field.}
\label{init}
\end{figure}

The NFFF model requires vector components of the magnetic field to perform the extrapolation. We have used the \texttt{hmi.sharp\_cea\_720s} magnetogram series with a temporal cadence of 12 minutes of  Helioseismic and Magnetic Imager \citep[HMI:][]{2012SoPh..275..207S}/SDO with the Lambert Cylindrical Equal Area (CEA) Projection \citep{2002A&A...395.1077C}. The cutout is originally 836 pixels in $x$- and 418 in $y$-directions. We performed the NFFF extrapolation for a rescaled size of 384 in $x$- and 192 in $y$- directions to reduce the computational cost. The physical extent of the computational box is $\approx $ 300 Mm in $x$- and $\approx $150 Mm in both $y$- and $z$-directions. We have compared the magnetic topology for the extrapolated field with the original and the rescaled resolution, which is found to be unchanged (not shown). We calculated the Pearson-r correlation coefficient for the extrapolated (${B}_{\rm{t}}$) and observed transverse (${b}_{\rm{t}}$) field and found it to be 0.95, suggesting a strong correlation between them. Panel (a) of Figure \ref{init} shows the observed magnetogram with $B_{\rm z}$ in the grayscale with the transverse positive and negative ﬁelds marked by blue and red arrows, respectively. Panel (b) shows a 2D map of the Lorentz force associated with the extrapolated field at the bottom boundary, with a $z$-component of it in grayscale and the transverse components marked by red and blue arrows. 

Next Figure \ref{extrap}(a) shows the overall field lines of AR 12824 obtained from the NFFF extrapolated magnetic field. We have plotted the field lines using the Visualization and Analysis Platform for Ocean, Atmosphere, and Solar Researchers (VAPOR) software \citep{clyne-2007,2019Atmos..10..488L}. Near the flaring region as shown in panel-(b), we find a set of twisted field lines, plotted in orange color, and confirm it as a flux rope from the estimation of the volumetric distribution of twist parameter ($Tw$), as shown in panel-(c). The twist parameter is a measure of the circulation of magnetic field lines about their axis. To calculate $Tw$, we have used the code of \citet{2016ApJ...818..148L} which computes the value by estimating $\vec{J} \cdot \vec{B}/B^{2}$ in a 3D  box. Further, a 3D magnetic null point is found near the flaring region. The field lines surrounding the spine and fan is plotted in white color. The null location is detected by using the method described in \citet{2020ApJ...892...44N}. The procedure is the realization of Dirac-delta function $\psi$ in a Cartesian grid which is found by constructing a Gaussian indicator in the form of 
\begin{equation}
     \psi(x,t) = \exp \Bigl[- \frac{(\vec{B}-\vec{B_{0}})^2}{d_{0}^2}\Bigl],
\end{equation}
\noindent where $\vec{B_{0}}=(B_{0}, B_{0}, B_{0})$, while $B_{0}$ and $d_{0}$ are small constants compared with the magnitude of $|B|$ that define an isovalue of $|B| $ and the spread of the Gaussian, respectively. For $B_{0} \approx  0$ and comparably small $d_{0}$, the function $\psi$ takes significant values only if $|B| \approx 0$. A 3D null is then within a small volume surrounded by an isosurface of $\psi$ with a suitably large value $\psi_{0}$. The method successfully identifies the null point locations in other examples as in  \citet{2021PhPl...28b4502N} near another flaring region. The null is pointed by the black arrow and its height is found to be $\approx$ 20 Mm.
%(x = 233 Mm, 298 pix, y = 84 Mm, y = 107 pix, z = 26 pix). 
In addition to the null point in the domain, a sheared arcade carrying a part of field lines of 3D null is also found lying below the fan of the 3D null, plotted in pink color. For further confirmation, we have plotted the distribution of squashing factor (Q) which represents the sharp change in the connectivity of the magnetic field lines. Panel (d) illustrates 3D null field lines overlaid with the squashing factor (Q) --- documenting its high values in the vicinity of the footpoints of the fan field lines.

\begin{figure}[!h]
\centering
\includegraphics[width=\textwidth]{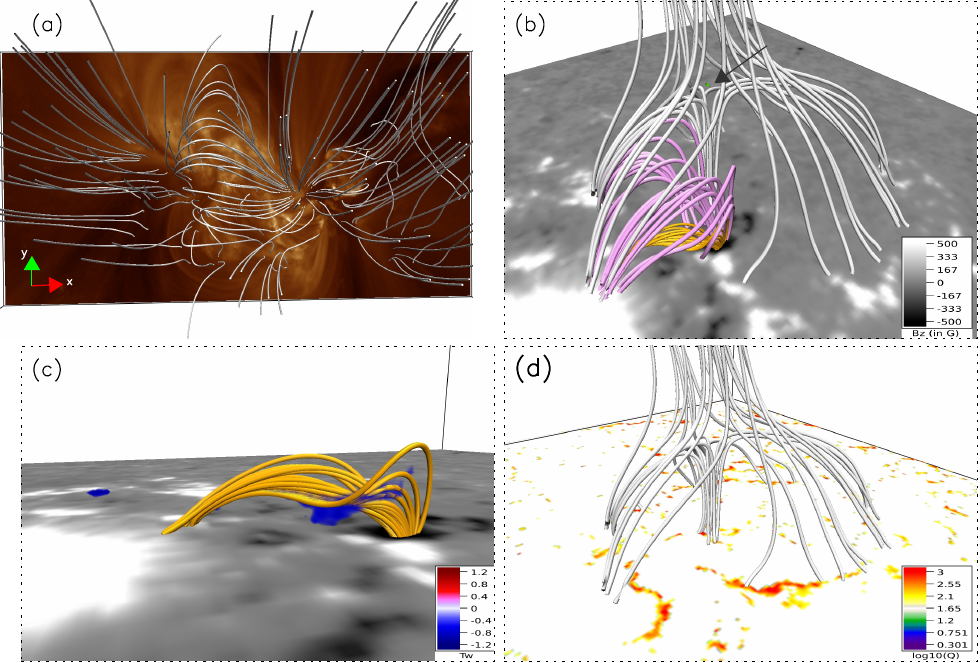}
\caption{The NFFF extrapolated field: (a) overall topology of the active region AR12884 plotted on 193 {\AA} of AIA. \textit{Panel (b)} magnetic field structure with a magnetic null point, plotted in \textit{white color}, the flux rope in \textit{orange color}, and the surrounding arcade in \textit{pink color}. The \textit{black arrow} and the \textit{cyan dot} mark the location of the null point. \textit{Panel (c)} highlights the flux rope structure possessing a maximum twist value of 1.2. The distribution of the squashing factor is plotted in \textit{panel (d)}. Notable is the presence of a high squashing factor near the footpoints of the fan field lines. }
\label{extrap}
\end{figure}

%--------Eulag----------------
\section{The Simulation Set-up for EULAG-MHD Model}
\label{numeric} 
To trace the footpoints on the ribbon structure, and to analyze the dynamics of the above topologies near the flaring region, we have carried out the simulation by numerically solving the incompressible Navier--Stokes MHD equations under the assumption of thermal homogeneity and perfect electrical conductivity \citep{sanjay2016, 2021SoPh..296...26K}. The MHD equations in the EULAG-MHD (Eulerian/Semi-Lagrangian Solver) model, in the dimensionless form, are
\begin{eqnarray}
\label{stokes}
& & \frac{\partial{\vec{v}}}{\partial t} 
+ \left({\vec{v}}\cdot\nabla \right) {\vec{ v}} =-\nabla p
+\left(\nabla\times{\vec{B}}\right) \times{\vec{B}}+\frac{\tau_{A}}{\tau_\nu}\nabla^2{\vec{v}},\\  
\label{incompress1}
& & \nabla\cdot{\vec{v}}=0, \\
\label{induction}
& & \frac{\partial{\vec{B}}}{\partial t}=\nabla\times({\vec{v}}\times{\vec{B}}), \\
\label{solenoid}
 & & \nabla\cdot{\vec{B}}=0, 
\label{e:mhd}
\end{eqnarray}
\noindent written in usual notations. The variables in the MHD equations are normalized as follows 
\begin{equation}
\label{norm}
{\vec{B}}\longrightarrow \frac{{\vec{B}}}{B_0},\quad{\vec{v}}\longrightarrow \frac{\vec{v}}{v_{A}},\quad
 L \longrightarrow \frac{L}{L_0},\quad t \longrightarrow \frac{t}{\tau_{A}},\quad
 p  \longrightarrow \frac{p}{\rho {v_{A}}^2}. 
\end{equation}
\noindent The constants $B_0$ and $L_0$ are generally arbitrary, but they can be fixed using the average magnetic field strength and size of the system. Here, $v_{A} \equiv B_0/\sqrt{4\pi\rho_0}$ is the Alfv\'{e}n speed and $\rho_0$ is the constant mass density. The constants $\tau_{A}$ and $\tau_\nu$ represent the Alfv\'{e}n transit time ($\tau_{A}=L_0/v_{A}$) and viscous dissipation time scale ($\tau_\nu= L_0^2/\nu$), respectively, with $\nu$ being the kinematic viscosity. Utilizing the discretized incompressibility constraint, the pressure perturbation, denoted by $p$,  satisfies an elliptic boundary-value problem on the discrete integral form of the momentum equation (Equation \ref{stokes}); cf. \citet{bhattacharyya+2010phpl} and the references therein. 

EULAG-MHD is based on the spatio-temporally second-order accurate non-oscillatory forward-in-time multidimensional positive definite advection transport algorithm MPDATA \citep{smolarkiewicz2006ijnmf}.  
Importantly, MPDATA has the proven dissipative property which, intermittently and adaptively, regularizes the under-resolved scales by simulating magnetic reconnections 
and mimicking the action of explicit subgrid-scale turbulence models \citep{2006JTurb...7...15M} in the spirit of
Implicit Large Eddy Simulations (ILES) \citep{grinstein2007book}. The residual numerical dissipation is ineffective elsewhere in the domain but not at the sites of reconnection such as the region with steep gradient in magnetic values or null points. Such ILESs conducted with the model have already been successfully utilized to simulate reconnections to understand their role in the coronal dynamics \citep{prasad+2017apj,prasad+2018apj,2019ApJ...875...10N, 2021PhPl...28b4502N,2022FrASS...939061K,2024ApJ...975..143N} in different solar transients like flares and jets. Details of the EULAG-MHD model can be found in \citet{smolarkiewicz&charbonneau2013jcoph}.

The NFFF extrapolated field is supplemented as the initial magnetic field and the initial velocity field is set to ${\vec{v}}=0$. At the bottom boundary, the $z$-components of ${\vec{B}}$ and  ${\vec{v}}$ are chosen to be fixed to their initial values throughout the simulation period (also termed as line-tied boundary condition) as the flux change during the flare is found to be minimum. The side and top boundaries are kept to their initial values. The density is kept to unity in the whole computational box. As stated earlier in Section \ref{extrap}, the simulation is initially driven by the non-zero Lorentz force associated with the extrapolated magnetic field, and the primary flow is generated by it. The resulting flow is however made incompressible following the Equation \ref{incompress1}, an assumption also adapted by \citet{dahlburg+1991apj, aulanier+2005aa}. Since our focus is to understand the onset of the flare through the topological changes, the assumption seems to be justifiable in the tenuous coronal medium. 

The computational domain in the simulation is of the same size to that of the extrapolation i.e. uniform grid resolution with 384 grid points in $x$-, and 192 in both $y$- and $z$- directions. The spatial unit length $\delta x$ is .0052 and the time step $\delta t$ is set to $1\times 10^{-3}$ while satisfying the CFL condition \citep{1967IBMJ...11..215C}. The dimensionless coefficient $\tau_A$/$\tau_\nu \approx 2 \times 10^{-4}$ in the simulation is roughly $\approx$15 times larger than its coronal value \citep{prasad+2018apj}. The parameter $\tau_A$/$\tau_\nu$ is controlled by the spatial resolution and the time step whilst satisfying the von Neumann stability criteria \citep{1947PCPS...43...50C}. The larger $\tau_{A}$/$\tau_\nu$, however, only expedites the evolution without an effect on the corresponding changes in the magnetic topology. The total simulation time is 4500 $\delta t$. To compare with the observational time, we multiplied the total simulation time by 15. Then, it corresponds to $\approx 33$ minutes of the observational time.

%----Result-----
\section{Simulation Results and Discussions}
\label{result}
\subsection{Onset of the Flare}
\begin{figure}
\centering
\includegraphics[width=\textwidth]{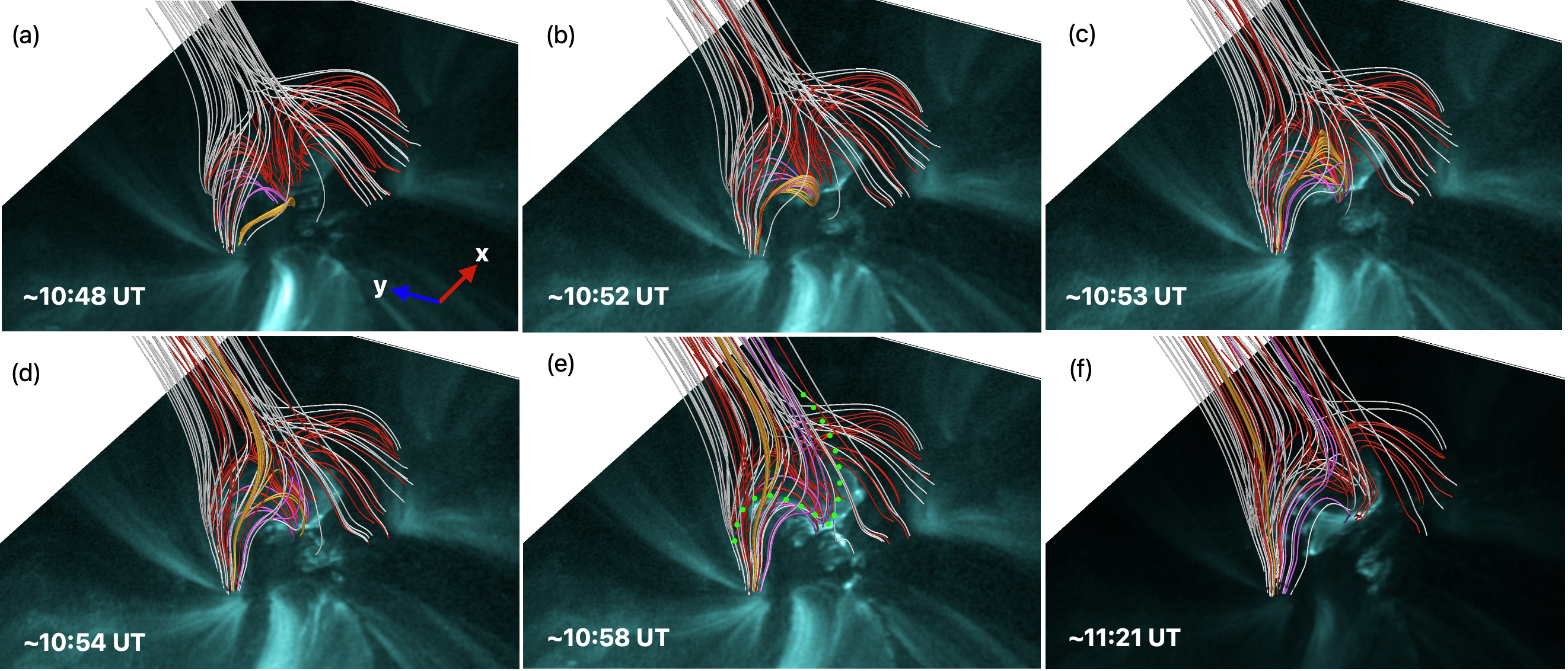}
\caption{\textit{Panels (a)-(f)} depict the evolution of magnetic field lines overplotted with 131 {\AA} on the bottom boundary during the whole flaring event. The change in connectivity seen in \textit{red-colored} field lines indicate toward a reconnection due to null point (\textit{panel (b)}). Simultaneously, the \textit{orange-colored} flux rope rises while untwisting itself and reconnect at the null point, shown in \textit{panels (b)-(d)}. The \textit{pink-colored} sheared arcade also reconnects at the null. The \textit{dotted line} in \textit{panel (e)} indicate a curved shape of the field lines that approximately resembles the brightenings seen in the background 131 images.  }
\label{evol131}
\end{figure}

\begin{figure}
\centering
\includegraphics[width=\textwidth]{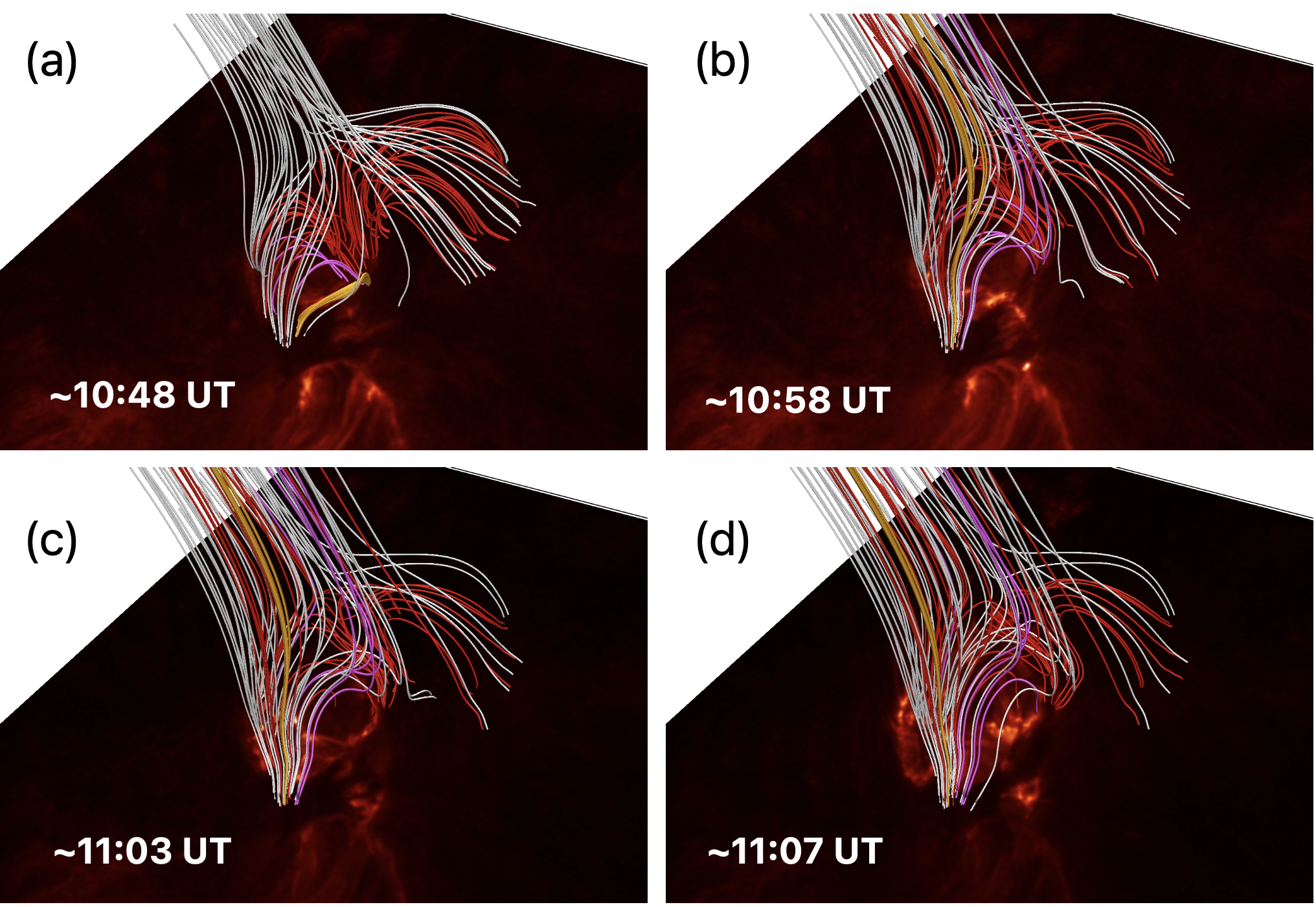}
\caption{Evolution of the same field lines as in Figure \ref{evol131} till the peak of the flare $t \approx $ 11:08 UT with 304 {\AA} images on the bottom boundary. Notable is the progress of the footpoints of the loops that coincide with the brightening and ribbons seen in 304 channel. These footpoints are considered as the sample field lines to trace the ribbon area later.} 
\label{evol304}
\end{figure}

\begin{figure}
\centering
\includegraphics[width=\textwidth]{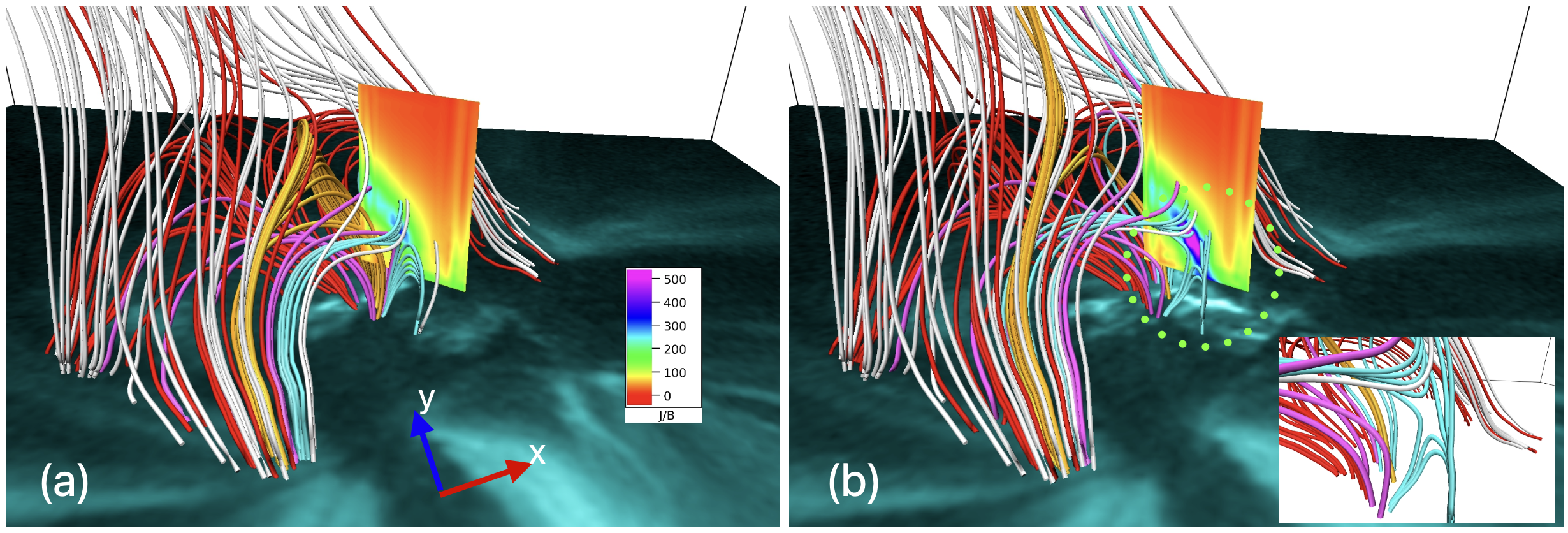}
\caption{\textit{Panels (a) and (b)} show the development of $|\vec{J}|/|\vec{B}|$ overlaid with 131 {\AA} images at time $t \approx $ 10:54 UT and $\approx$ 10:56 UT, respectively. The figure documents the formation of a current sheet beneath the flux rope. The \textit{dotted green circle} in \textit{panel (b)} highlights the X-shaped geometry of field lines near the current sheet. The \textit{inset} in the \textit{panel (b)} provides its zoomed-in view.}
\label{jbyb}
\end{figure}

First, we have shown the evolution of the field lines of the null point skeleton (white and red colors), the flux rope (orange) and the sheared arcade (pink) found in the vicinity of the flaring region. We have compared the dynamics developed in different magnetic field configurations with AIA 131 and 304 channels. In Figure \ref{evol131}, we have overplotted the sample field lines with 131 channel on the background. From panels (a)-(f), we have shown the evolution of the field lines tracing over the flaring region from $t \approx$ 10:48 UT to $t \approx$ 11:21 UT covering the whole flaring period. Panel (a) denotes the initial state of the simulation, whereas panel (f) corresponds to end of the flare. Panels (b) to (e) denote snapshots before the peak of the flare time period. Initially, the red-colored field lines are situated under the dome of the 3D null (panel (a)) and, with time, the field lines come out of the dome (panel (b)). This clearly indicates a change in connectivity of the red field lines, suggesting the onset of the reconnection at the 3D null. These reconnections repeat in time and can contribute to the brightening observed around the inner spine of the 3D null. Subsequent dynamics documents the rise of the flux rope toward the 3D null and its reconnection at the null (panels (c) and (d)). Similar rise and reconnection are also observed for the sheared arcade. Such rise and reconneciton lead to development of a curved magnetic structure, marked by a dotted green line in panel (e). The structure has some similarity with the observed bright structures. In Figure \ref{evol304}, we have plotted the evolution of the field lines overlaid with 304 {\AA}, highlighting the overlying of their footpoints in the brightened region near the peak of the flaring period. 
The opening of these different field lines along the outer spine of the 3D null  may indicate a direction to the plasma flow. Furthermore, we have seen the formation of a current sheet, as evident from the enhancement in $|\vec{J}|/|\vec{B}|$ in Figure \ref{jbyb}. 
The inset in the panel (b) shows an X-shaped geometry of field lines, also marked by a dotted green circle, in the vicinity of the current sheet. As a result, magnetic reconnection is found to initiate there, which may also contribute to the observed flare brightenings in 131 {\AA}.

\begin{figure}
\centering
\includegraphics[width=.47\textwidth]{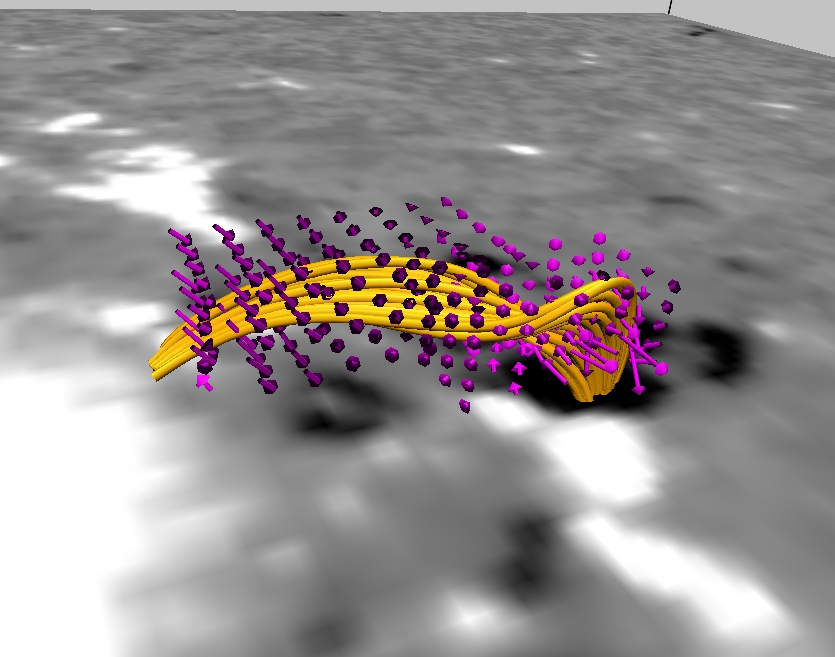}
\includegraphics[width=.47\textwidth]{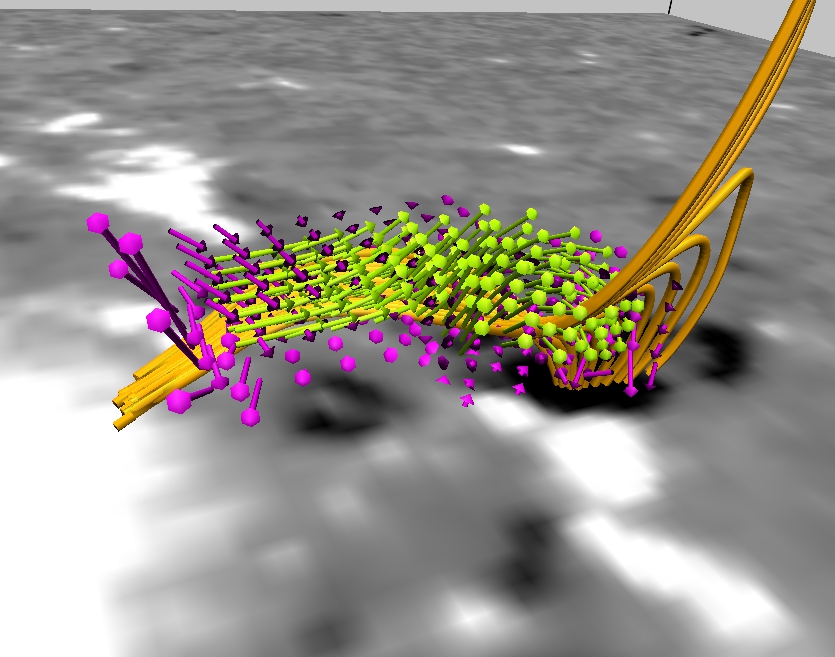}
\caption{\textit{Left panel} depicts the Lorentz force streamlines in \textit{pink} surrounding the flux rope at the initial time, where plasma flow is absent. The \textit{grayscale} on the background is for the $B_{z}$ component. The \textit{right panel} shows an interim state where the eventual development of plasma flow can be seen, plotted in \textit{green color}. The surrounding Lorentz force helps in unwinding the flux rope first and then the ambient plasma flow adds to the dynamics of the flux rope.} 
\label{lforce}
\end{figure}

To further examine the dynamics near the flux rope, in Figure \ref{lforce}, we have plotted the Lorentz force (marked by pink arrows) and the plasma flow (denoted by green arrows) at the initial time and few minutes after that in the vicinity of the rope. The Lorentz force seems to be initiating the untwisting of the flux rope (left panel), where the plasma flow is absent owing to the initial set-up in the numerical model. The right panel shows a development of plasma flow in the region that governs the dynamics of the rope field lines. Here, the untwisting of the field lines can be attributed to the complementary forcing due to the Lorentz force and the plasma flow. The unwinding of the flux rope is a telltale sign of the contribution toward loss of plasma materials inside it \citep{2019MNRAS.490.3679W,2023ApJ...949....2Z}. Moreover, in the right panel, the rope field lines appear to rise toward the 3D null to reconnect --- another contributory factor to the material loss.  Interestingly, recent works by
\citet{2023ApJ...951L..35C,2023ApJ...959...67C} has attributed the outer null point reconnection for deciding the nature of the flare to be eruptive or confined in a breakout type model \citep{1999ApJ...510..485A,2008ApJ...680..740D}. According to them, the poloidal flux in a rope may be disrupted due to the null point reconnection occurring above it, and hence destructing its structure too. However, in our case, the inherent $\vec{J}\times \vec{B}$ and the developed plasma flow facilitate the rise and untwisting of the flux rope, that ultimately lead to the eruption.


\subsection{Calculation of Reconnection Flux and Reconnection Flux Rate}
\subsubsection{Using \citet{2007ApJ...659..758Q}'s method}
\begin{figure}
   \includegraphics[width=1\linewidth]{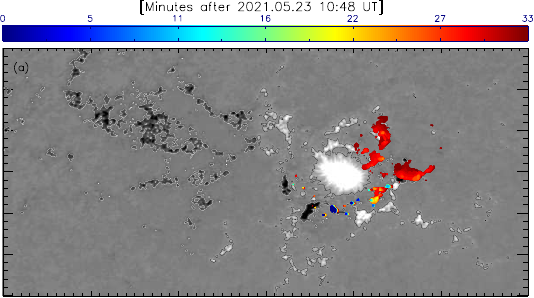}
    \includegraphics[width=1\linewidth]{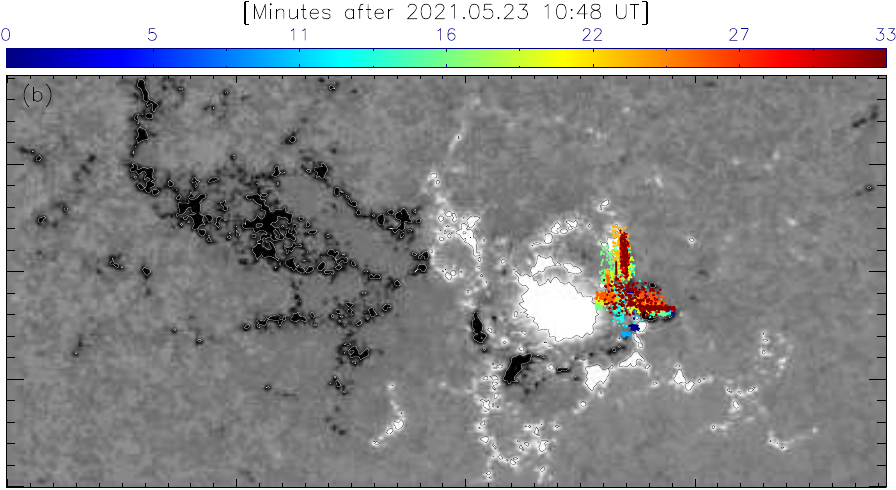}
\caption{\textit{Panel (a)}: time evolution of flare ribbons plotted over the $B_{z}$ component of the magnetic field. The \textit{black and white contours} depict the positive and negative polarities respectively. The whole observation duration here corresponds to the whole simulation period. \textit{Panel (b)}: time evolution of ribbon area obtained by tracing the sample field lines from the simulation at certain time steps. Notable is the overall agreement with the time evolution of flare ribbons obtained in \textit{panel (a)}.}
\label{ribbon}
\end{figure}

To understand the reconnection process through flare ribbon dynamics, we have estimated both accumulated reconnection flux and reconnection flux rate using flare ribbons. For the purpose, we have adapted the method by \citet{2002ApJ...565.1335Q, 2007ApJ...659..758Q} which quantifies the participation of magnetic fluxes in the flare. Following Equation (1) of \citet{2007ApJ...659..758Q} in the hindsight of the CSHKP model, the reconnection rate is expressed as 
\begin{equation}
\label{rrate}
\frac{\partial \phi_{r}}{\partial t} = \frac{\partial}{\partial t} \int B_{c} {\rm{d}}S_{c} = \frac{\partial}{\partial t} \int B_{l} {\rm{d}}S_{l},
\end{equation}
\noindent which provides a relationship between the rate of change of the integration of coronal magnetic field ($B_{c}$) over the reconnection area $S_{c}$ to the normal magnetic field component ($B_{l}$) in the lower atmosphere passing through the ribbon area ($S_{l}$). The algorithm in \citet{2007ApJ...659..758Q} uses the flare ribbons as a proxy to trace the footpoints of the magnetic field lines, reconnecting at higher heights. Then, it remaps those brightened pixels to the HMI magnetogram to extract the +ve ($\phi_{r}^{+}$) and -ve fluxes 
($\phi_{r}^{-}$). For identifying the ribbon pixels used in the calculation of the reconnection flux and the corresponding rate, the method uses 1600 channel of AIA/SDO and $B_{z}$ component of the magnetogram from HMI/SDO, shown in Figure \ref{ribbon}. 
At every timestamp, the brightened ribbon pixels are selected when the intensity of the pixel is more than six times the median intensity and this is determined from the averaged 6-minute time cadence maps before the eruption of the event. The threshold corrects for the saturation or projection effects \citep{2022ApJ...934..103H}. 
We extracted the ribbons pixels which is shown in panel (a) of Figure \ref{ribbon}. They are overplotted on the HMI/SDO magnetogram. 

\subsubsection{Using simulated data}
\begin{figure}
\centering
\includegraphics[width=.5\textwidth]{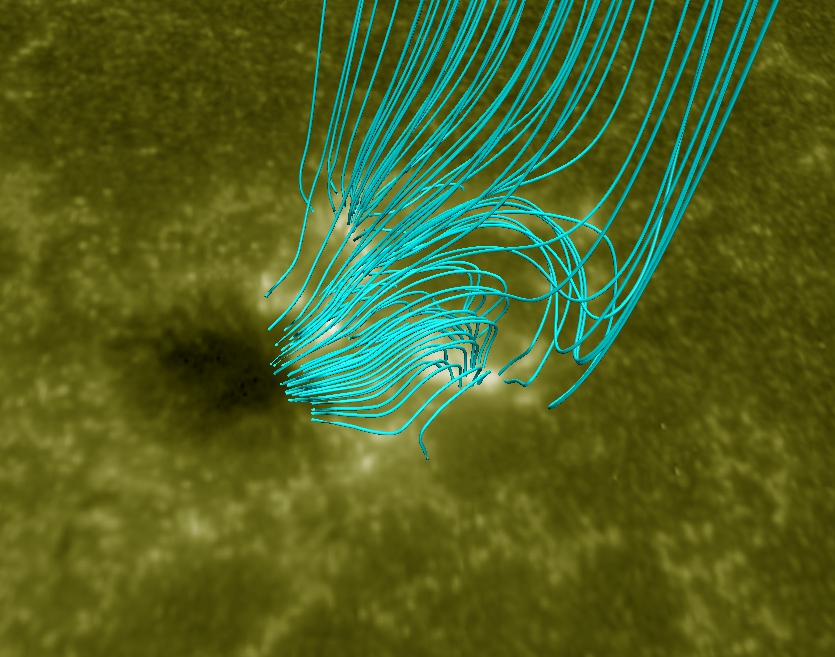}
\caption{The sample field lines are plotted on the ribbons observed in the 1600 channel. The footpoints of these field lines are traced to calculate the reconnection flux and flux rates.} 
\label{ribbonfl}
\end{figure}

\begin{figure}
   \includegraphics[width=\linewidth]{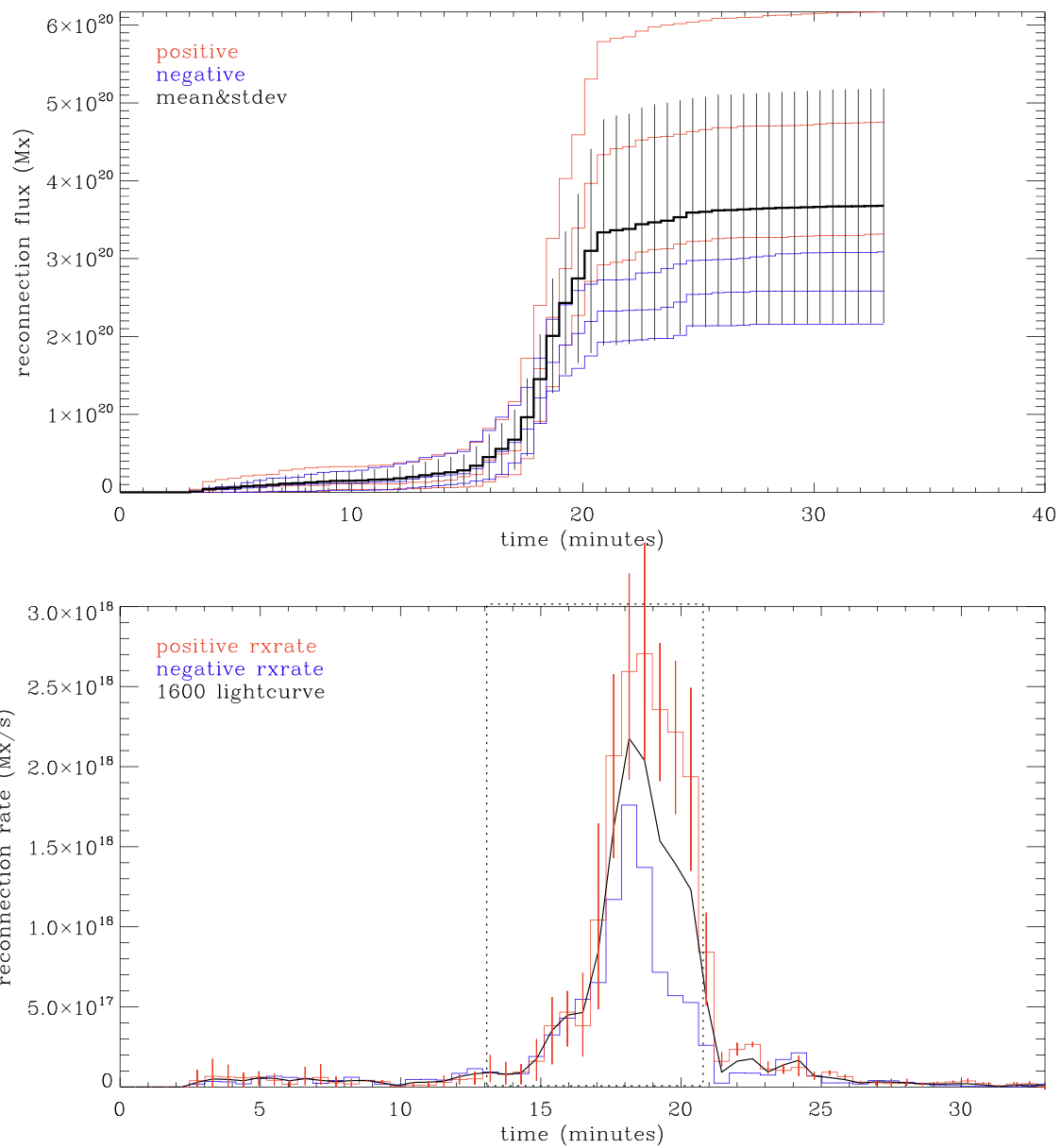}
\caption{\textit{Top panel}: the reconnection flux during the whole flare time, for both positive and negative polarities passing through the ribbon area which was traced by 1600 channel. \textit{Bottom panel}: the reconnection flux rate derived from the above for both the polarities. The \textit{dashed box} denotes the time window for which the reconnection flux and reconnection flux rates are calculated in the simulation. }
\label{reconflux_obs}
\end{figure}

\begin{figure}
\centering
   \includegraphics[width=.85\linewidth]{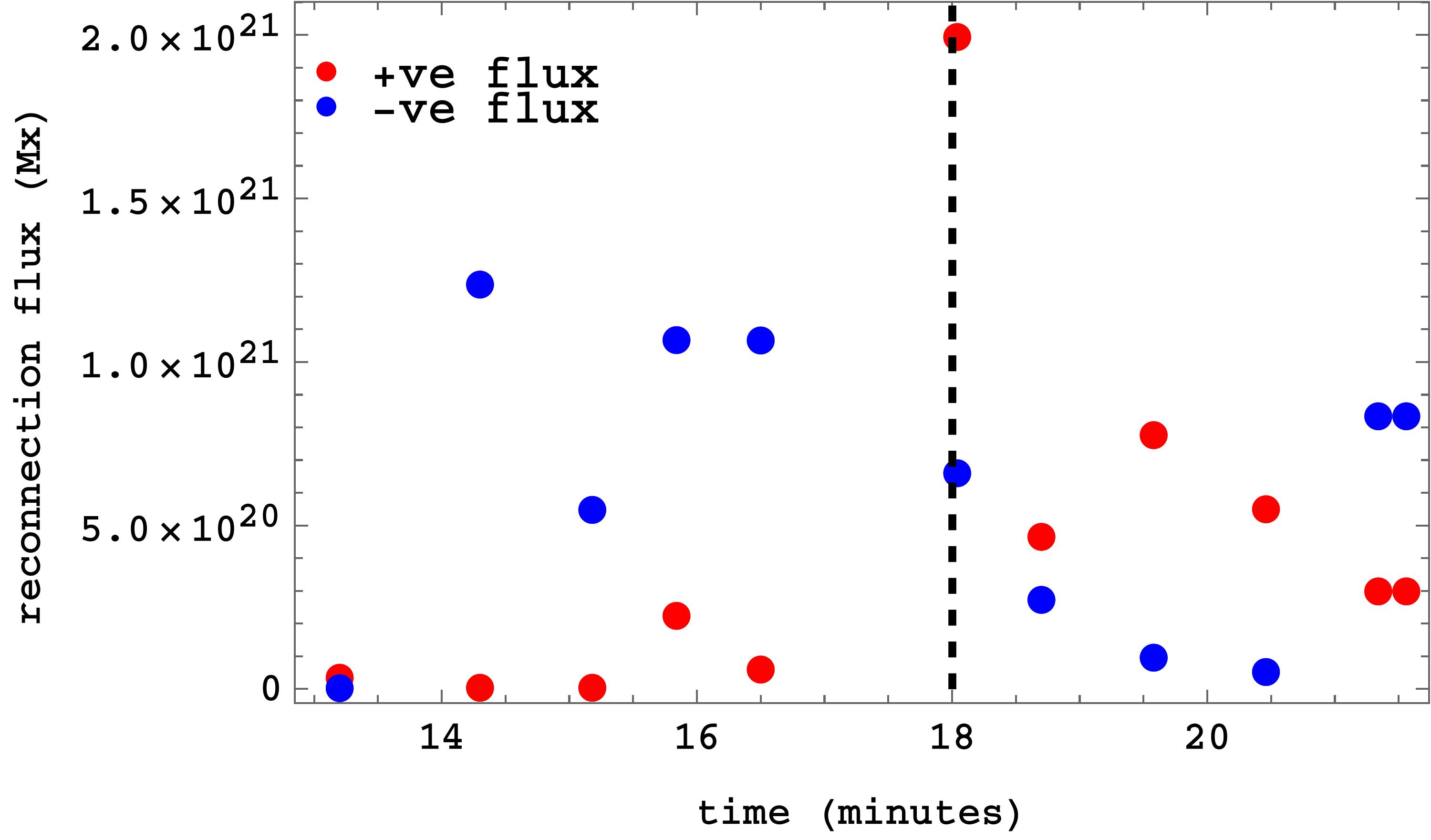}
    \includegraphics[width=.81\linewidth]{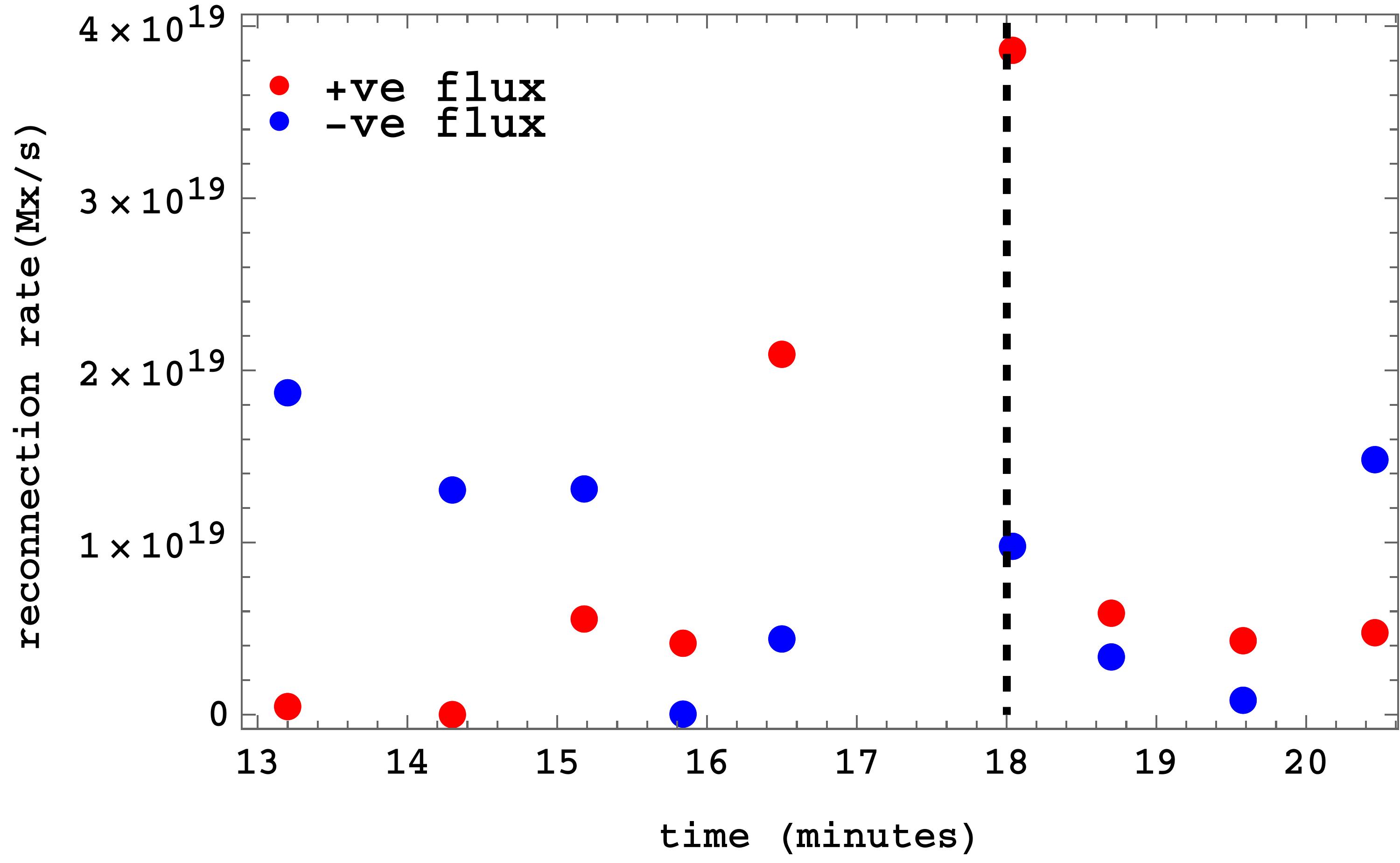}
\caption{The \textit{top and bottom panels} represent the reconnection flux and the reconnection flux rate derived from the simulation within the ribbon area. Noteworthy is the similar pattern in comparison to the observation flux rate however 10 times higher than the observational values. The \textit{dashed line} in \textit{black color} highlights the peak time of the flare.}
\label{reconflux_sim}
\end{figure}

For the calculation of the reconnection flux and reconnection flux rate in the simulation, we have followed the similar steps of \citet{2007ApJ...659..758Q}'s method, which basically rely on the sites involving ribbon pixels. First of all, we have identified the primary reconnection site which in this case is the 3D null point. Then we have traced the footpoints of the field lines of inner spine and fan of the null over the flare ribbons. 
This choice is based on the understanding that the charged particles accelerated by the reconnection at the null are expected to follow these field lines and, therefore, the corresponding footpoints are anticipated to co-locate with the flare ribbons. 
Relatively, an observation-based study by  \citet{2002ApJ...566..528I,2005A&A...441..353N} have used similar approach to identify the footpoints of reconnecting loops. In this study, they have used cusp-shaped loops in decay phase of a flare observed in \texttt{Yohkoh} Soft X-Ray Telescope, and Transition Region And Coronal Explorer (TRACE) data to extract the footpoints. To demonstrate the chosen field lines, in Figure \ref{ribbonfl}, we have plotted the field lines over the ribbons in 1600 {\AA} channel at the peak time of the flare.
\noindent We have purposely avoided tracing the field lines with the same seed points during reconnection, as new field lines may reconnect there. During the evolution, we have extracted the footpoints of the sample field lines over each time step of the simulation that is passing through the flare ribbon areas and separated the +ve and -ve magnetic fluxes. As stated earlier in the boundary conditions in Section \ref{numeric}, the $B_{z}$ component is not evolving throughout the simulation time, however $B_{x}$ and $B_{y}$ are changing over each time step. This is important as the connectivity changes each time and is essential for the tracking of ribbon pixels during the simulation. The justification behind the calculations is that the relation in Equation \ref{rrate} is valid for a line-tied boundary condition \citep{2007ApJ...659..758Q}. After extracting the footpoints, we then calculated the reconnection flux, $\phi_{+/-}$ for the corresponding pixels using the following formula:
\begin{equation}
\label{rrate-sim}
\phi_{+/-} =  \sum_{i=1}^{N}A_{i}{{B_{z_{(+/-)}}}_{i} }, 
\end{equation}
\noindent where, N is the total number of pixels, A is the area of the pixel and $B_{z}$ is the normal magnetic field.  We have plotted those footpoints location in the panel (b) of Figure \ref{ribbon}. The agreement can be seen near the kernel of the flaring region. Next, in Figure \ref{reconflux_obs} we have plotted the estimated reconnection flux and reconnection flux rates from the observations for the considered simulation period. The dashed box in the bottom panel of Figure \ref{reconflux_obs} highlights the range for the start of impulsive and the decay phases of the flare. We compared same estimations with the simulated ones. The simulated reconnection flux and reconnection flux rates are plotted in Figure \ref{reconflux_sim}. In the top panel, we can see the peak of the flux attaining an order of $10^{21}$ which is one order higher than the observed value. The plot in the bottom panel shows the flux rates, which is also one order higher than the observed rate. The extra one order may be attributed to the higher $\tau_A$/$\tau_\nu \approx 2 \times 10^{-4}$ in the simulation.  However, in both the panels, the increasing profile of flux and flux rate of +ve pixels is found to be similar to that of observed +ve flux cases, whereas the simulated profile for -ve fluxes has some deviation from the observed one. 

\subsection{Induced Electric Field Near 3D-null Reconnection Region}
\begin{figure}[!h]
\centering
\includegraphics[width=\textwidth]{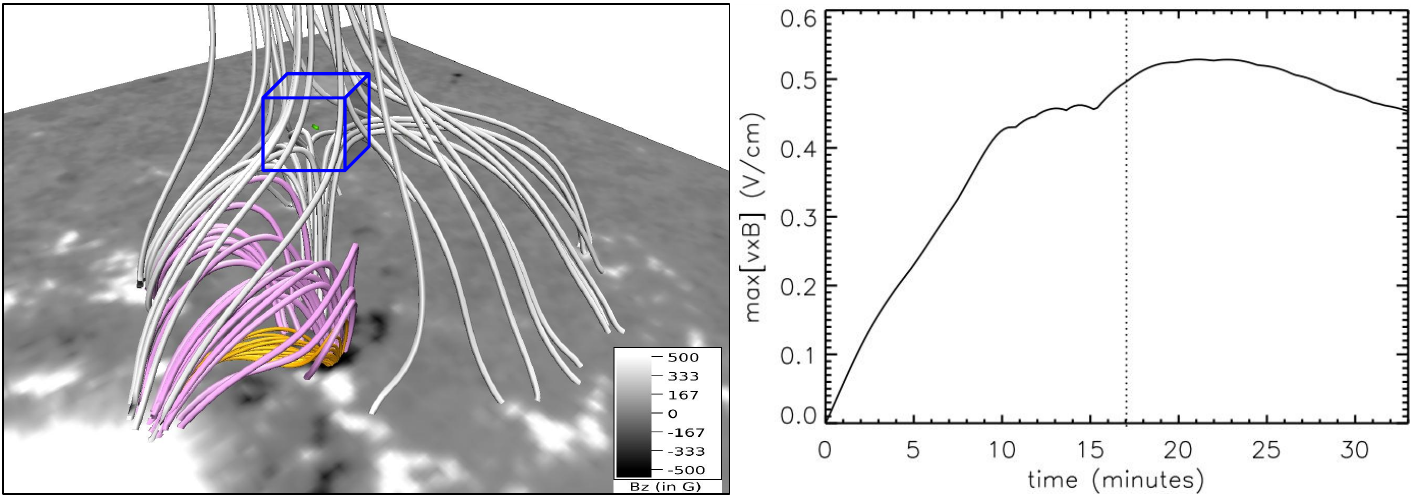}
\caption{Estimation of $|\vec{v}\times \vec{B}|$ close to the null location showing the profile of reconnection rate: \textit{blue box} in \textit{left panel} denotes the volume under consideration used for the estimation. The \textit{right panel} shows the temporal variation of  $|\vec{v}\times \vec{B}|$ in that volume. The maximum value reached in this flare is $\approx $.52 V cm$^{-1}$.}
\label{vcb}
\end{figure}

To understand the variation of reconnection electric field as another way to estimate the reconnection rate during the reconnection near the 3D null point, we have calculated the corresponding $|\vec{v}\times \vec{B}|$ parameter surrounding the null point. We have considered a volume of $25\times 25\times 25$ pixels in each direction keeping the null point in the center of this small cube. The left panel of Figure \ref{vcb} marks the volume of interest. The right panel shows the profile of the  $|\vec{v}\times \vec{B}|$ for the whole simulation time. We find the maximum value of the electric field reached during the flare is $\approx $.52 V cm$^{-1}$, which is slightly less than the values obtained by  \citet{1998SoPh..178..125P}, where they found the values to be 1-3 V cm$^{-1}$ in case of solar flares. Moreover, \citet{1986NASCP2442..469K} have suggested the values of electric field intensity within a concrete flare to be 1–1.5 V cm$^{-1}$ from the velocity of flare ribbons , which provides a
sufficiently large voltage along the reconnection line. Further, our estimated value is much lesser than that of \citet{1983SoPh...83...83F} in case of post-flare coronal loops which was found to be 170 V cm$^{-1}$.

\subsection{Evolution of the Current Density Near the Ribbons}
\begin{figure}
\centering
\includegraphics[width=.6\textwidth]{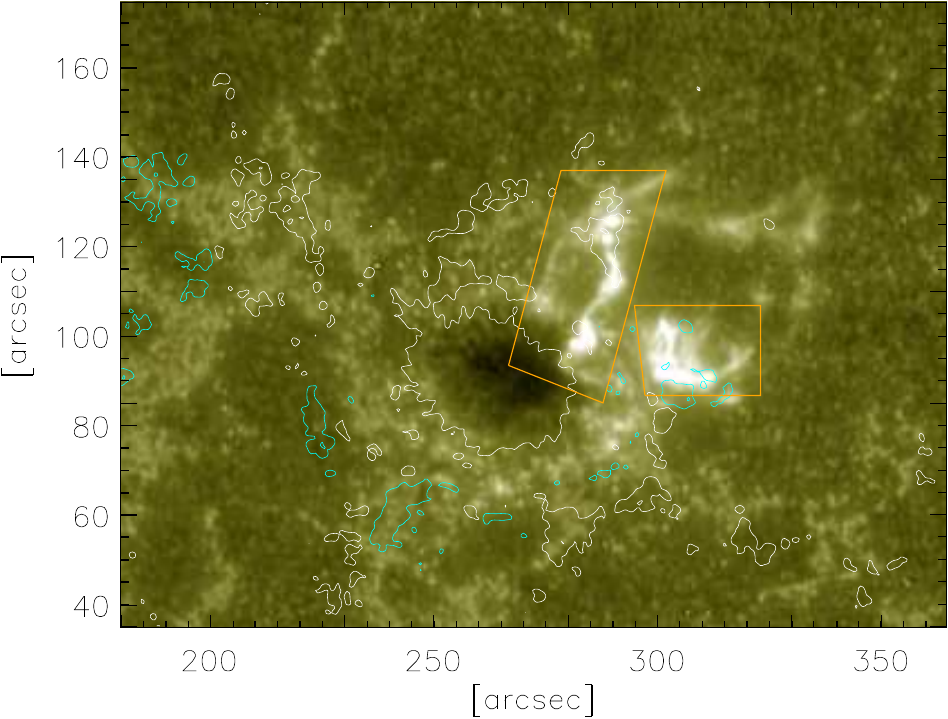}
\includegraphics[width=.95\textwidth]{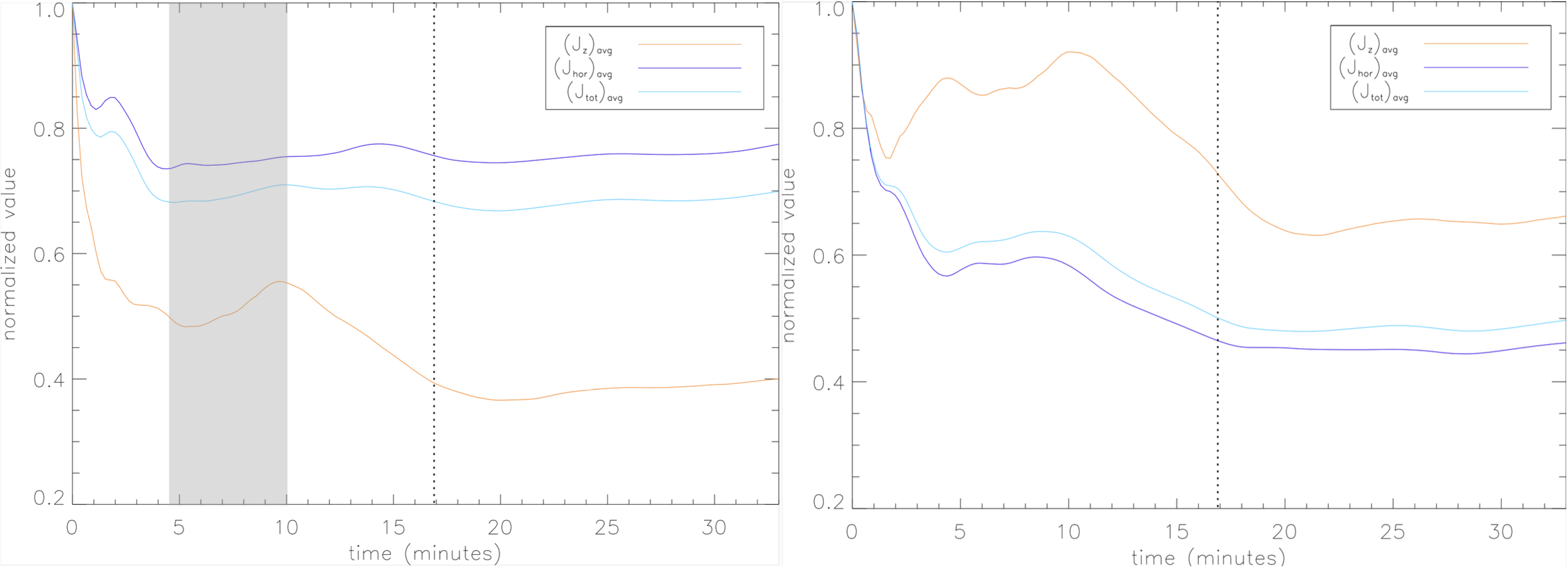}
\caption{\textit{Top Panel}: selection of ribbon areas for which different components of currents are calculated. The background image is from 1600 channel with $B_{z} \in ({-100, 100})$ G contours on it. Evolution of averaged vertical current density ($J_{z}$), the horizontal current 
density ($J_{\rm hor}$), and the total current density ($J$) obtained through simulation for the pixels under the \textit{orange boxes} for the photospheric (\textit{left-bottom panel}) and near-chromospheric (\textit{right-bottom panel}) layers. Notable is the increase in the $J_{z}$ before the peak of the flare, also marked by the \textit{shadowed region}.} 
\label{current_evol}
\end{figure}

Next, we have investigated the evolution of current density near the flaring region during the simulation. Ribbon areas marked in orange color boxes of the top panel of Figure \ref{current_evol}. We have plotted the time profile of the different current components, depicted in the bottom panels, such as the averaged vertical current density ($J_{z}$), the horizontal current density ($J_{\rm hor}$), and the total current density ($J$) in blue, cyan and orange colors respectively for the flare ribbons locations which correspond to photospheric (left-bottom panel) and near-chromospheric  (right-bottom-panel) layers. The dashed line indicates the flare peak time. Interesting is a significant increase in $J_{z}$ (marked by a shadowed region) during the pre-flare/near-impulsive phase of the flare in both layers. In their study of electric currents within the flare ribbons of AR 11158, \citet{2014ApJ...788...60J} observed a comparable pattern (as illustrated in Figure 5 of their work) showing an increase in direct current during the impulsive phase of the flare.
The only distinction noted was that the upward trend appeared a few minutes earlier in our analysis. This is again in congruent with the pattern in the reconnection flux in Figure \ref{reconflux_obs}. The increase in the currents and later decay of them suggest the dissipation of the magnetic energy via reconnection at the null point and its lower spine region. This is again in favor of the formation of the brightened pixels as ribbons. 
 
\begin{figure}
\centering
\includegraphics[width=.495\textwidth]{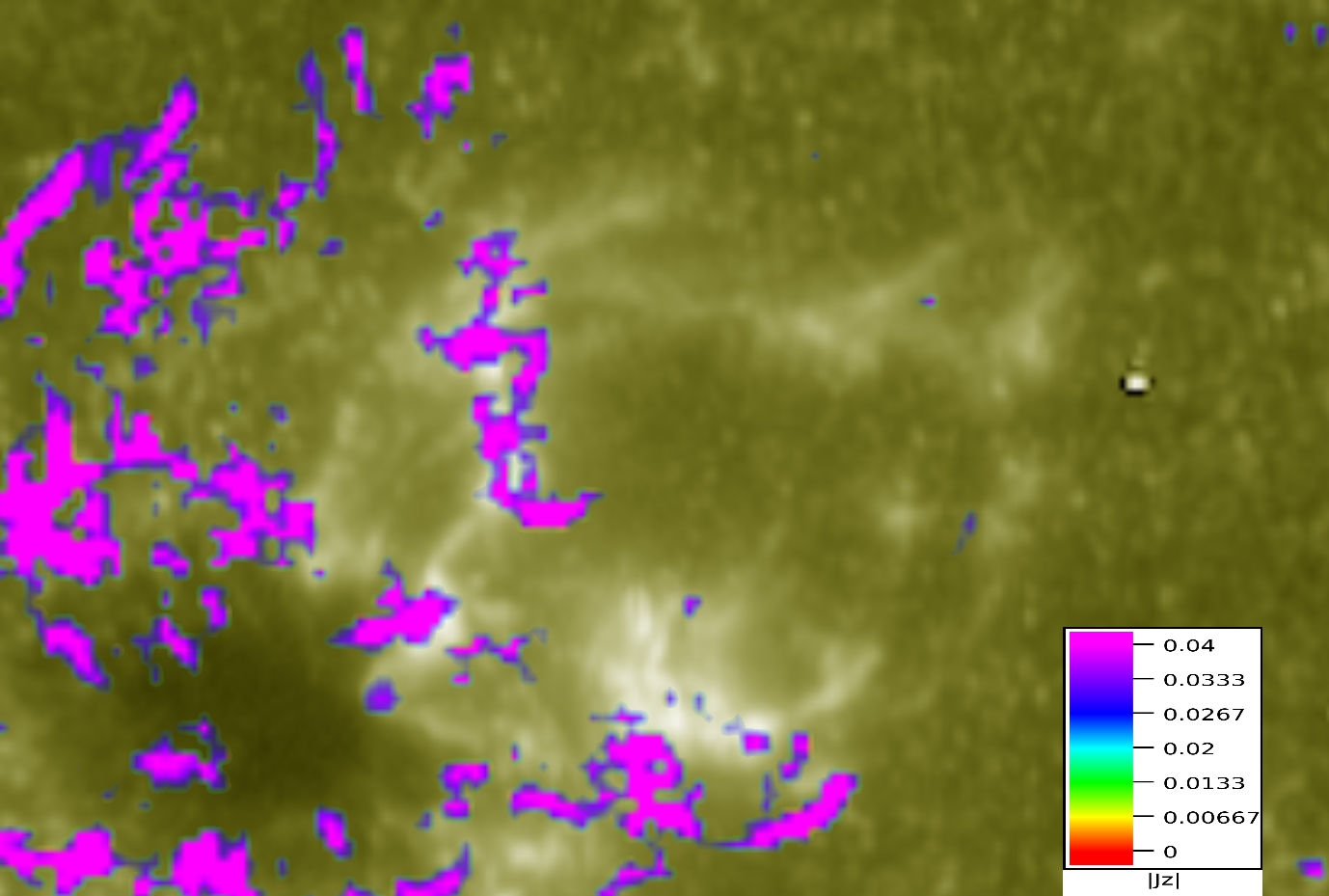}
\caption{Distribution of $J_{z}$ overplotted on the ribbons. The background is a snapshot from 1600 channel at peak time of the flare $\approx$ 11:08 UT.}
\label{snap}
\end{figure}

To further aid the argument of the heating due to the currents, in Figure \ref{snap}, we have overlaid the distribution of the vertical current density $J_{z}$ on the ribbons computed from the simulation. The pattern of their distributions on the near-photospheric layer matches reasonably well with that of the ribbons. In their study, \citet{2016A&A...591A.141J} also compares the traces of the photospheric currents to the evolution of flare ribbons in an eruptive flare and find the similar morphology and evolution there through a magnetofrictional method. Importantly, the high values of currents are found to coincide with the footpoints of the flux rope and the field lines of the 3D null. The topological evolution from Figure \ref{evol131} and the evolution of $J_{z}$ near the ribbons in Figure \ref{current_evol} indicate the increase in the non-potentiality first in the sample field lines and then followed by the dissipation with the continuous magnetic reconnection near the null point. These complementary actions of reconnection through the null point and the occurrence of high $J_{z}$ are the precursors for the formation of ribbons. The flux rope loses its non-potentiality due to Lorentz force and rises as a result of null point reconnection and makes a passage for the plasma materials to escape later. 

%----Summary------------
\section{Summary}
      \label{conclusion} 
In this paper, we have studied the triggering mechanism and subsequent ribbon formation by the underlying magnetic reconnection of an M-class flare in the AR 12824. For this purpose, we have performed a data-constrained magnetohydrodynamics simulation. We have extrapolated the active region's magnetic field using the non-force-free-field extrapolation model. The active region exhibits a complex magnetic field topology. In the magnetic field extrapolation, we find a magnetic null point and a flux rope near the flaring kernel. Importantly, the interaction of these structures is found to play a crucial role in initiating the eruptive event and provides insights into the onset mechanism. To explore the dynamics of these structures during the flare, we have studied the simulated dynamics and compared them to the contemporary observations. 

We analyzed the reconnection scenario leading to the flare onset and its further development, and the effect of reconnection on the flare ribbon dynamics. Null point magnetic topology is found near the flaring region exhibiting a non-parallel ribbon structure which helps in the onset of the reconnection. The reconnection continues there, and due to the high-pressure gradient, the flux rope lying below the null point rises and participates in the reconnection, slightly mimicking a jet-type eruption. Eventually, a current sheet is also formed below the flux rope and a further X-type reconnection is followed there aiding to the brightenings near the flaring region. The footpoint imprints obtained from the simulation are catching up with the flare ribbons deduced from the observations. The estimation of reconnection flux and flux rates from the MHD simulation follows a similar pattern to that of observations, however, differ one order quantitatively when compared to the observationally computed values. We have calculated the flare-induced electric field surrounding the null point in the corona and found the maximum value reaching 0.52 V cm$^{-1}$ during the peak of the flare which is slightly low compared to many earlier works. Lastly, the current enhancement in the flare ribbons during the impulsive phase of flare indicates the successive reconnections and formation of flare ribbons due to the bombardment of charged particles from the upper reconnection site.

 Summarily, the work with data-constrained MHD simulation in tandem with observations presents the onset mechanism of a complex flare triggered by magnetic reconnection, analysis of the flare ribbon dynamics, and the driver for the ribbon formation in a complicated magnetic field configuration. Apart from the topological evolution, the simulation captures the flare ribbon dynamics which is well compared with the observations. Albeit the boundary conditions in the simulation model, the work succeeds in not only explaining the complex magnetic drivers for the flare and corresponding ribbons but also puts forward a scenario not seemingly following the standard flare model. However, the limits of simulation are to deliver a more realistic evolution of the fluxes, their role in driving the flare and particularly ribbons, the development of non-potentiality surrounding the flaring region and its impact on the eruption process, and the formation of flux rope in a complex magnetic region which are still unclear. In this direction, a future attempt will focus on improving the boundary condition for the studies of flare ribbon dynamics. 

\begin{acks}
 We thank Prof. Jiong Qiu for the technical help in the reconnection flux calculation model. We acknowledge the Bladerunner cluster at the Center for Space Plasma and Aeronomic Research at the University of Alabama in Huntsville. Data and images are courtesy of NASA/SDO and the HMI and AIA science teams. SDO/HMI is a joint effort of many teams and individuals to whom we are greatly indebted for providing the data.
\end{acks}

\noindent
\begin{authorcontribution}
 S.S.N. and Q.H. constructed the goal of the work. S.S.N. carried out data acquisition, prepared plots, performed simulation, and wrote the main draft of the paper. W.H. helped in preparing Figure 6 and 8. Q.H., W.H. and S.K. contributed to the overall discussion of the work. S.K. and R.B. contributed to the interpretation of the simulation results. All the authors did a careful proofreading of the draft.
\end{authorcontribution}
 
\begin{fundinginformation}
S.S.N. and Q.H. acknowledge supports 
from NSF-AGS-1954503, 
NASA-LWS-\\
80NSSC21K0003, 
and 80NSSC21K1671 
grants.
S.K. would like to acknowledge the support from the 
ANRF-SERB project No. SUR/2022/000569 
and Patna University project No. RDC\\/MRP/07. 
\end{fundinginformation}

\begin{dataavailability}
The data that support the findings of this study are available from
the corresponding author upon request.
\end{dataavailability}

\begin{conflict}
The authors declare that they have no conflicts of interest.
\end{conflict}
%%% BIBLIOGRAPHY %%%%%%%%%%%%%%%%%%%%%%%%%%%%%%%%%%%%%%%%%%%%%%%%%%%%%%%%%%%

\bibliographystyle{spr-mp-sola}
    
\bibliography{reference}  

\end{article} 

\end{document}